# Accurate determination of the absolute $^3$He/$^4$He ratio of a synthesized helium standard gas (Helium Standard of Japan, HESJ): Towards revision of the atmospheric $^3$He/$^4$He ratio

**Kenji Mishima[1*], Hirochika Sumino[2], Takahito Yamada[3], Sei Ieki[3], Naoki Nagakura[3], Hidetoshi Otono[4], Hideyuki Oide[3†]**

[1] High Energy Accelerator Research Organization (KEK).

[2] Dept. of Basic Sci., The University of Tokyo.

[3] Grad. School of Sci., The University of Tokyo.

[4] RCAPP, Kyushu University.

[*]Corresponding author: Kenji Mishima, KEK, Japan (kenji.mishima@kek.jp)

[†]Current address: INFN Sezione di Genova

**Key Points:**

- The absolute $^3$He/$^4$He ratio of the helium standard of Japan (HESJ) was determined with accuracy of 0.40%.
- The atmospheric $^3$He/$^4$He ratio was determined as $1.340 \pm 0.006$ ppm.

**Abstract**

The helium standard of Japan, referred to as HESJ, is an inter-laboratory standard for the $^3$He/$^4$He ratio. While the ratio of $^3$He and $^4$He of the HESJ was previously determined by a relative comparison to atmospheric helium, the absolute value of the $^3$He/$^4$He ratio of the HESJ has not been directly determined yet. Therefore, it relies on the early measurements of that of atmospheric helium. The accuracy of the absolute $^3$He/$^4$He ratios of the atmosphere and other working standards including HESJ is crucial in some applications of helium isotopes, such as tritium-$^3$He dating, surface-exposure age determination based on cosmogenic $^3$He, and the accurate measurement of the neutron lifetime. In this work, new control samples of helium gases with $^3$He/$^4$He ratios of 14, 28, and 42 ppm were fabricated with accuracy of 0.25-0.38% using a gas-handling system for a neutron lifetime experiment at Japan Proton Accelerator Research Complex (J-PARC). The relative $^3$He/$^4$He ratios of these samples and the HESJ were measured using a magnetic-sector-type, single-focusing, noble gas mass spectrometer with a double collector system. As a result, the absolute $^3$He/$^4$He ratio of the HESJ was determined as $27.36 \pm$

0.11 ppm. The atmospheric $^3$He/$^4$He ratio was determined as 1.340 ± 0.006 ppm, based on this work.

# 1 Introduction

The $^3$He/$^4$He ratios of terrestrial samples varies by more than three orders of magnitude because the primordial $^3$He/$^4$He ratio of (1.7–4.6) × 10$^{-4}$ (Porcelli & Ballentine, 2002) has been diluted by radiogenic $^4$He, produced by decay of U- and Th-series elements; the degree of dilution depends on the $^3$He/(U+Th) ratio of each geochemical reservoir, such as the mantle and crust. Owing to this fact, the $^3$He/$^4$He ratio is a powerful tracer in geochemistry and cosmochemistry (Ozima and Podosek, 2002). Atmospheric helium, with an absolute $^3$He/$^4$He ratio of 1.3-1.4× 10$^{-6}$ (Clarke et al., 1976; Mamyrin et al., 1970; Meija et al., 2016; Sano et al., 2013) has been used as a common reference sample in order to calibrate the $^3$He/$^4$He ratio measurements with noble-gas mass spectrometers. It has been indicated that the relatively low $^3$He/$^4$He ratio and low fraction of atmospheric helium leads to practical difficulties in measurement by statistics of $^3$He counts or effect of impurities. Thus, in several cases, research groups of noble-gas laboratories create their common local working standard samples with a relatively high $^3$He/$^4$He ratio, produced either from a natural gas sample with a relatively a high $^3$He/$^4$He ratio, or from a mixture of isotopically-pure $^3$He and $^4$He.

The helium standard of Japan, hereafter referred to as HESJ, which falls into the latter category, was originally created by four noble-gas laboratories in Japan, and is now distributed worldwide as an inter-laboratory standard (Matsuda et al., 2002). However, the $^3$He/$^4$He ratio of the HESJ ($R_{HESJ}$) has not directly been measured, but only been determined relatively to that of atmospheric helium, thus its accuracy relies on early determinations of the absolute $^3$He/$^4$He ratio of atmospheric helium ($R_a$). Though the $R_a$ value had been claimed not to be temporally or spatially constant (Sano et al., 1988, 2008), it was not supported by later studies (Lupton & Evans, 2013; Mabry et al., 2015). Since $^3$He/$^4$He ratio is generally used to compare relative contributions of primordial and radiogenic helium sources in each geochemical reservoir, the absolute $^3$He/$^4$He ratio of the atmospheric helium or that of the HESJ is not necessarily required.

Nevertheless, knowing the absolute $^3$He/$^4$He ratio is critical in certain applications of helium isotopes. Examples of such cases are tritium-$^3$He dating (Schlosser, 1992; Takaoka & Mizutani, 1987; Visser et al., 2014), cosmogenic $^3$He-based surface exposure age determination (Niedermann, 2002), and an experimental project to measure the neutron lifetime using a pulsed neutron source at Japan Proton Accelerator Research Complex (J-PARC) (Arimoto et al., 2015; Nagakura et al., 2016).

The tritium-$^3$He dating is a method to determine age of groundwater or seawater since it has been isolated from the atmosphere. Tritium in the atmosphere is produced by nuclear reaction of the air and cosmic rays or by nuclear weapons. It decays into $^3$He with the half-life of 12.33(6) years (Firestone et al., 1996). Thus, simultaneous tritium and $^3$He measurements make it possible to estimate the time since when a certain water sample has been isolated from the atmosphere underground or in the deep ocean. The amount of $^3$He, usually determined by $^3$He/$^4$He ratio measurement, directly affects the tritium-$^3$He age because the age is a function of $^3$He to tritium ratio.

The cosmogenic $^3$He-based surface exposure age determination is a method to estimate time for helium-retentive minerals to be exposed on the surface of the earth. The exposure age can be determined through amount of $^3$He produced by spallation of mineral-forming nuclei by cosmic rays if helium diffusivity in the minerals is enough low to neglect helium escape during geological timescales of interest. For instance, production rates of $^3$He in a olivine and pyroxene phenocrysts at sea level were estimated as 117–138 atoms/g/year (Niedermann, 2002). Thus, an absolute $^3$He/$^4$He calibrator is required to calculate a sample age of exposure.

A neutron decays into a proton, an electron, and an anti-neutrino with a lifetime of 880.2 ± 1.0 sec (Patrignani et al., 2017). The lifetime of the neutron is an important constant in the Big Bang nucleosynthesis, which controls the abundance of primordial elements in the universe. In the neutron lifetime experiment at J-PARC mentioned above, the decay volume of the detector, referred to as the time-projection chamber (TPC), is filled with a mixture of gas of $^3$He, $^4$He, and $CO_2$ (Arimoto et al., 2015; Nagakura et al., 2016). The incident flux of the cold neutron is measured by counting the rate of the $^3$He(n,p)$^3$H reaction in the TPC. Here, the $^3$He density in the detector needs to be known accurately in order to determine the neutron flux. In the experiment, a gas-handling system is used in order to control the $^3$He number density with an uncertainty of approximately 0.3%. In this study, the control samples of multiple $^3$He/$^4$He ratios were fabricated using this system. The relative $^3$He/$^4$He ratios to the HESJ of these samples were measured by a modified VG5400, which is a magnetic-sector-type, single-focusing noble-gas mass spectrometer with a double collector system at Department of Basic Science, of the University of Tokyo (Sumino et al., 2001). The results can contribute to determination of the absolute $^3$He/$^4$He value of the HESJ, and that of atmospheric helium as well.

## 2  $^3$He/$^4$He control samples

The control samples were fabricated to have the same level of $^3$He/$^4$He ratio to that of HESJ (approximately 28 ppm). In this study, three control samples of 14, 28, and 42 ppm were produced by mixing diluted isopure $^3$He and $^4$He gases, by using a gas expansion method for the accurate mixture. The gas expansion method is a way to inject a small amount of gas accurately by using the diffusion of two well-known volumes. In this section, the procedures of gas fabrication are described.

### 2.1 The gas handling system

A schematic view of the gas-handling system is shown in Figure 1. The gas-handling system consists of 1/4- and 3/8-inch stainless tubes and bellows seal valves (Swagelok SS6BK) connected by Swagelok joints. Four sectors, $V_0$–$V_3$, are defined. The system is equipped with a turbo molecular pump (TMP) and gas-sampling bottles.

The stainless tubes themselves were defined as $V_0$, $V_1$, and $V_2$, whose volumes were approximately 43, 95, and 14 cm$^3$, respectively. The buffer bottle $V_3$, with a size of ⌀210.3 mm × 635 mm and a volume of 22 × 10$^3$ cm$^3$ was used to dilute $^3$He gas. The handling of gases, i.e., the introduction, dilution, and extraction of the $^3$He and $^4$He gases, was performed via $V_1$. Two absolute pressure gauges were used to measure induced and diffused gas pressures; a piezoresistive transducer (Mensor CPG2500) and a Baratron manometer (MKS 690A11TRA), connected to $V_0$ and $V_2$, respectively. The piezoresistive transducer had two gauges with different full scales of 120 kPa and 35 kPa with accuracies of 6 Pa (or 0.01% in the range of 60–120 kPa) and 3.5 Pa, respectively. The full scale of the Baratron gauge was 1.33 kPa, with an accuracy of 0.05% of the reading, and its temperature coefficient was 4 ppm/K of the full scale and 20 ppm/K of the reading, respectively. The sensor of the Baratron gauge was kept at 45 °C during operation. The temperature of the gas handling system was monitored by two platinum resistance thermometer sensors (PT100) attached at the front of the gas panel and the buffer bottle, where the accuracy of temperatures was 65 mK.

Isopure gases of $^3$He and $^4$He (ISOTEC), connected to $V_1$, were used to fabricate the control samples. The contamination of $^4$He in the isopure $^3$He gas was less than 0.05%, and the contamination of $^3$He in the isopure $^4$He gas was 0.5 ± 0.2 ppb, according to their specification. The contaminations in the gases, such as $H_2O$ or $N_2$, measured by a quadrupole mass spectrometer, was less than $3\times10^{-4}$ in total, which is enough smaller than required sensitivity of this work, and neglected.

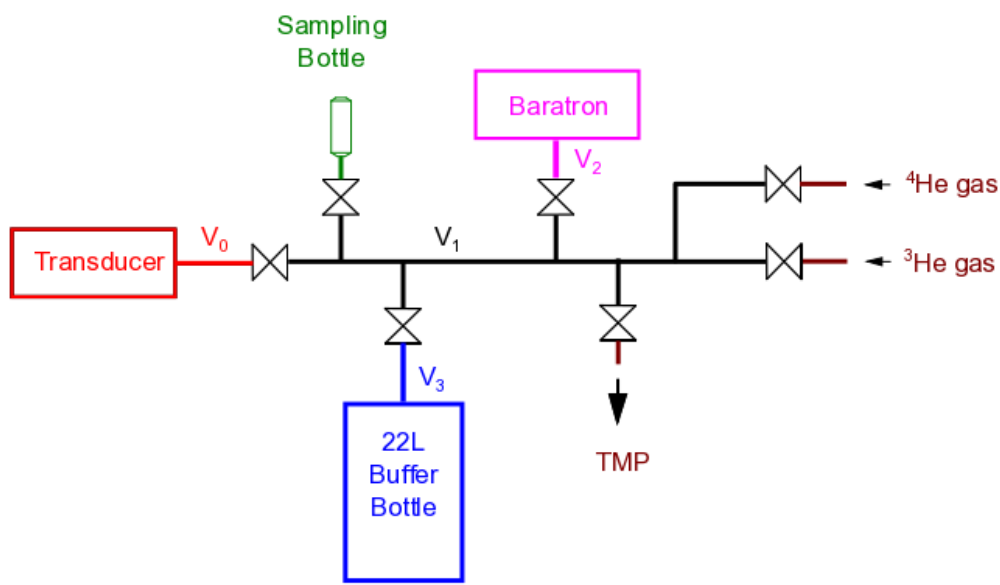

**Figure 1.** Schematic of the gas handling system. Four sectors ($V_0$–$V_3$) are defined as volumes divided by valves. The system is evacuated by a turbo molecular pump (TMP).

2.2 Determination of volume ratios

In order to produce 14-42 ppm of the gas mixture, the $^3$He gas has to be diluted with high accuracy. This was achieved by evaluating the proper corrections to the ideal gas assumption in the determination of the relative volume ratios between different sectors. The bare volume ratio was measured by comparing the change of the pressure of helium gas in expansion from one volume to the both, by assuming ideal gas conditions. In reality, the helium gas does not behave as an ideal gas, and a residual correction was applied by using the second virial coefficient of 11.83(3) cm$^3$/mol (Kell et al., 1978). The correction was 0.12% with uncertainty of $3\times10^{-6}$. The

Baratron was operated at 45 °C, which is ~20 K higher than gas volumes, and it is known that the measurement had a bias at the relatively low-pressure region due to the thermal transpiration effect (Setina, 1999). However, the size of this bias was found to be negligible (approximately 6 × $10^{-5}$) for the operated pressure of approximately 600 Pa. Note that the effective volume change due to the operation temperature of 45 °C is expected to 3 × $10^{-6}$ because the high temperature region in the Baratron is small (~1 $cm^3$) comparing 22 liter buffer volume. The volume measurements have been done following procedure; first, commercial helium gas was filled by an initial volume. Then, the gas was released to the other volume. After 1 min of waiting to be stable the gas condition, the released pressure was measured by the pressure gauges.

Table 1 summarizes the results of volume ratios in the various combinations of initial and final volumes. Numbers in the bracket show a standard deviation of uncertainty. The values of ratios $A$, $B$, and $C$ were directly obtained from the measurements, while the value of $D$ was calculated using $B$ and $C$. The uncertainties were calculated as the sum of the uncertainties of all pressure measurement with a full correlation among them.

**Table 1.** Volume ratios measured by the gas expansion method

| Name | Volumes | Ratio | Relative percent |
|---|---|---|---|
| A | $(V_0)/(V_0+V_1)$ | 0.30783(6) | 0.02% |
| B | $(V_0+V_1)/(V_0+V_1+V_2)$ | 0.905(2) | 0.22% |
| C | $(V_0+V_1)/(V_0+V_1+V_2+V_3)$ | 0.006135(5) | 0.07% |
| D | $(V_0+V_1)/(V_0+V_1+V_3)$ | 0.006139(5) | 0.07% |

2.3 Production of the control samples

Three control samples of 14, 28, and 42 ppm of the absolute $^3$He/$^4$He ratios were created. These ratios were chosen to be close to that of the HESJ of approximately 27 ppm. In order to achieve several tens ppm of the mixture ratio, the dilution of $^3$He by $^4$He was performed twice using the 22-L buffer volume ($V_2$). The procedure was as follows:

1. $^3$He was filled in $V_0 + V_1$ to be the required pressure ($P_1$:2–4 kPa) by slowly opening the valve. After evacuating $V_1$, then $V_0$ was diffused to $V_0 \rightarrow V_0 + V_1 + V_3$.

2. After evacuating $^3$He in $V_0+V_1$, $^4$He was filled in $V_0+V_1$ ($P_2$: approximately 100 kPa).

3. $^3$He in $V_3$ and $^4$He in $V_0+V_1$ were mixed in $V_0+ V_1+V_3$.

4. The mixed gas in $V_0$ was left and $V_1+V_3$ were evacuated. Following the evacuation, $V_0$ was released to $V_0 \rightarrow V_0+ V_1+V_3$.

5. After evacuating the mixed gas in $V_0+V_1$, $^4$He was filled in $V_0+ V_1$ ($P_3$: approximately 60 kPa)

6. $^3$He in $V_3$ and $^4$He in $V_0+V_1$ were mixed in $V_0+ V_1+V_3$.

7. The fabricated gas was sampled in a sampling bottle.

The $^3$He/$^4$He ratios of the fabricated samples can be determined by using the volume ratios of $A$ and $D$ in Table 1 and the initial gas pressures of $P_1$–$P_3$ as

$$\frac{^3He}{^4He} = \frac{P_1 D A^2 (1-D)^3}{P_3 + DA(1-D)^2 P_2}. \tag{1}$$

The $^3$He/$^4$He ratios of the control samples were adjusted to be 14, 28, and 42 ppm by controlling the $^3$He pressure of $P_1$, which are listed in Table 2. Numbers in the bracket show a standard deviation of uncertainties.

Table 2. $^3$He/$^4$He ratios of the fabricated and isopure $^4$He samples

| Sample number | $^3$He/$^4$He ratio [ppm] | Relative percent |
|---|---|---|
| I | 14.01(5) | 0.38% |
| II | 28.05(8) | 0.30% |
| III | 42.01(11) | 0.25% |
| Isopure $^4$He | 0.005(2) | 40% |

The maximum temperature difference before and after the expansion was 0.90 K. The effects of the change of the temperature-dependent volume ratios were corrected linearly using the measured temperature values. The bellow seal valve changed its volume of 0.2 cm$^3$ by opening/closing the valve; however, it was negligibly small compared to the 22-L volume.

It may have taken a significant amount of the time to complete the diffusion in step 3 and step 6 of the above procedure. The diffuse time was determined in-situ by sampling the gas by varying the time after finishing step 3, as shown by the result in Figure 2. It was observed that the $^3$He/$^4$He ratio ramps up with time and saturates. The data, as a function of elapsed time $t$ after the beginning of mixing, is fitted by

$$f(t) = f_{sat}\left\{1 - \exp\left(\frac{-t}{\tau_d}\right)\right\}, \tag{2}$$

where $f_{sat}$ is the $^3$He/$^4$He ratio at the saturation, $\tau_d$ is the time constant of the diffusion. Note that we budgeted the uncertainty of valve operation time had been 5 sec. The value of $\tau_d$ was determined as 70 ± 3 s. The diffusing time in steps 3 and 6 was taken more than 30 min. Therefore, the gas mixture sample was sufficiently uniform in the procedures.

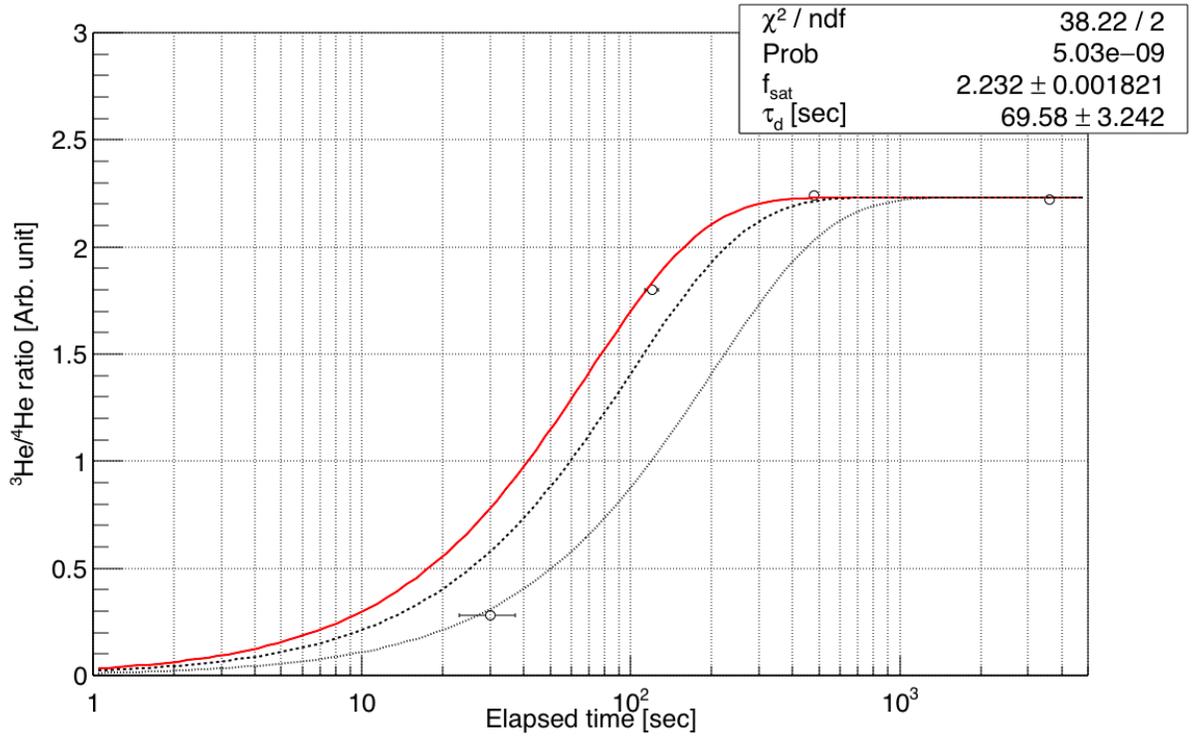

**Figure 2.** $^3$He/$^4$He ratios with different mixing times. The $^3$He/$^4$He ratios were determined with mass spectrometry. The solid red line shows a fit of eq. (2) to the data. The fitted parameters are shown in the box. The black dash, and dotted line show with different $\tau_d$ value of 100 and 200 sec, respectively.

## 3  $^3$He/$^4$He mass spectrometry

The $^3$He/$^4$He ratio of the HESJ was determined by control samples fabricated for calibration, as described in the previous section. A magnetic-sector-type single-focusing noble gas mass spectrometer (MS) with a double collector system at the Department of Basic Science, of the University of Tokyo (Sumino et al., 2001) was used for the measurements. The mass spectrometer has an ion counting detector for $^3$He, which is composed of an electron multiplier, an amplifier, a discriminator and a counter to count number of amplified signals of $^3$He ions entering into the multiplier, and a Faraday cup equipped with an amplifier and $10^{10}$ ohm feedback register for $^4$He; thus, $^3$He and $^4$He can be measured simultaneously using a fixed magnetic field. The period of the measurement for each sample was 400 s. Before ion counting, the magnetic field was scanned and set to an optimized field where the peak centers of $^3$He and $^4$He coincide which is the least sensitive to magnetic field fluctuations.

The measurements were performed 5 times for each of control samples. In order to suppress the time fluctuations of the MS outputs, each sample and the HESJ were measured alternately. The HESJ and the control sample gases were buffered in 1.5-L containers and introduced into the MS with a pressure in the range of 3–7 × 10$^{-6}$ Pa following the chemical purifying processes. In this MS, 2.4% decrease in measured $^3$He/$^4$He ratio was observed with helium partial pressure exceeding $6.4 \times 10^{-4}$ Pa (Sumino et al., 2001). However the pressure range of helium admitted to the MS during this study is far lower than the limit, and given that there is any pressure effect depending on the pressure difference between the analyses of HESJ and control samples, the maximum pressure difference of $4 \times 10^{-6}$ Pa in this work would result in only 0.015% difference in $^3$He/$^4$He ratio. Note that we did not measure any blank samples during the measurement because constant backgrounds would be canceled in this experimental procedure.

A typical time spectrum of the $^3$He/$^4$He value of the HESJ is shown in

Figure **3**. The vertical axis shows the ratio of signals of the $^3$He ion detector to the $^4$He Faraday cup with errors calculated by adding the errors of both detectors in a quadrature. A typical $^3$He ion detector count rate and $^4$He Faraday cup current during the HESJ measurements were $(700 \pm 26)$ cps and $(8000 \pm 3)$ fA, respectively, where the $^3$He count error, equivalent to statistical error, is dominant in $^3$He/$^4$He error. The relative effect of dead time due to the pileup is expected to be ~$10^{-5}$ because the pulse shape of the ion detector is ~10 ns. The horizontal axis shows the elapsed time following the injection of the sample gas. Each point corresponds to a measurement duration of 40 s.

Figure 4 shows a time spectrum of the isopure $^4$He measured 5 times longer than the normal measurement. A significant increase of $5.2 (4) \times 10^{-7}$/s, corresponds to the $^3$He/$^4$He ratio of $1.0 (1) \times 10^{-5}$ ppm/s, was observed. This is known as the memory effect, and it is due to release of the implanted gas of previous measurements into the source and collector slits and inner wall of the flight tube of MS. However, the effect was negligible because the increase of the $^3$He/$^4$He ratio in the duration of the measurement of 1000 s is about $1.0 (1) \times 10^{-2}$ ppm, which was 1000 times less than those of the HESJ or the control samples. Thus, we ignored the effect, and the $^3$He/$^4$He ratios of the measurements were determined by the time average of all points.

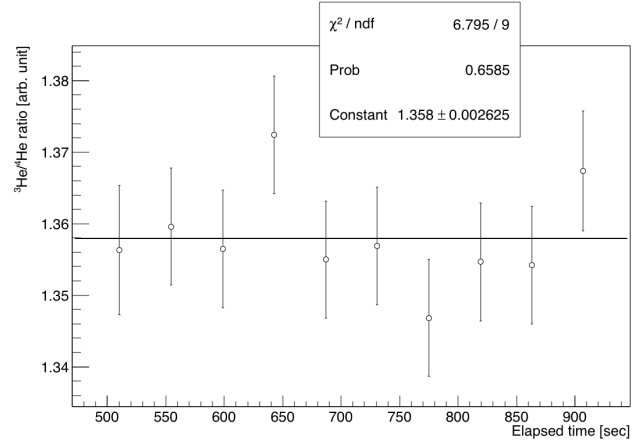

**Figure 3.** Typical time spectrum of $^3$He/$^4$He measurement for the HESJ. Each point corresponds to a measurement duration of 40 s. The error bars are dominated by the statistics of $^3$He counts.

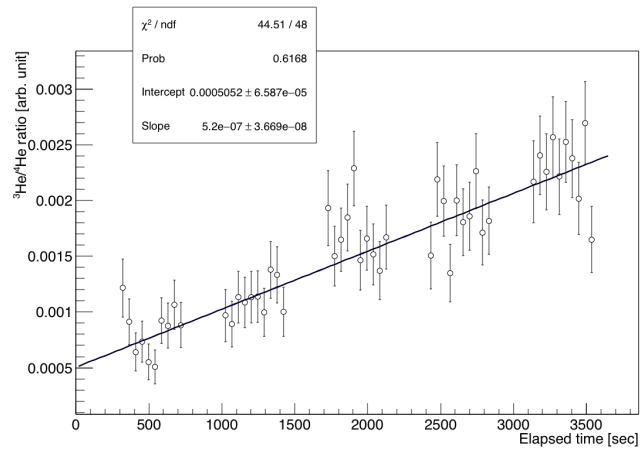

**Figure 4.** Time spectrum of the $^3$He/$^4$He measurement for the isopure $^4$He. An increase of 5.2 (4) ×10$^{-7}$/s in $^3$He/$^4$He ratios with time would result from memory effect (see text).

The determined $^3$He/$^4$He ratios for all measurement are shown in Figure 5 and then for each sample are shown in Figure 6 for the HESJ (a), control samples (b–d), and the isopure $^4$He (e) with the fittings by constants. The errors shown with points are statistical errors of 1 σ.

The data in Figure 6 did not agree within the statistic error, where 68% of them should be in 1 σ. Since the peak position was calibrated before each run to compensate a possible position shift, a slight deviation of the magnetic field from that corresponding to the peak centers of $^3$He and $^4$He is unlikely to be the reason of the scattering of $^3$He/$^4$He ratios. Moreover, magnetic field and temperature of the magnet of the MS were monitored during the measurements but no significant correlation of them with the $^3$He/$^4$He ratios ($r < 0.3$) was observed. As another possibility of the origin of the scattering $^3$He/$^4$He, we suspect instability of the ion source that could change transmission of $^3$He through a slit at the front of the ion counter, whilst its effect on $^4$He would be negligible because a slit at the front of the Faraday cup is about three times wider than that of the ion counter. This effect would change the collection rate of $^3$He but not $^4$He, resulting in fluctuated $^3$He/$^4$He ratios beyond internal error of each measurement.

In that case, a simple fitting with only statistic error gave us too small error than the reality. Thus, we evaluated the errors of averaged value by taking account the scattering of data by multiplying a scale factor defined as

$$S = \sqrt{(\chi^2/\text{n.d.f.})} \ . \qquad (3)$$

The scale factor increased the fitting error to be the reduced $\chi^2$ ($\chi^2$ divided by the number of degree of freedom, $\chi^2$/n.d.f.) as unity which means that it is almost equivalent to determine the error by the scattering of data. See introduction of Patrignani et al. (2017) for the detail of the method. In case for the HESJ, the $\chi^2$/n.d.f. was 546/20. Thus, the error was multiplied by a factor of 5.2. All data and fit results in Figure 6 are in a supporting information.

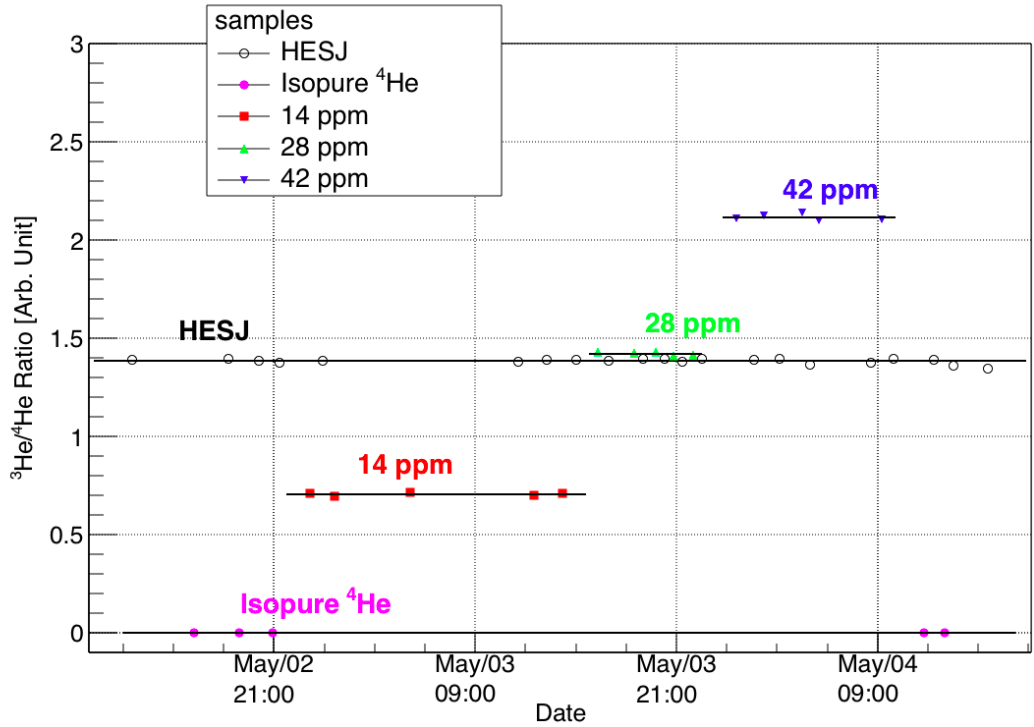

**Figure 5.** Time spectrum of measured ratios of $^3$He/$^4$He for the HESJ, control samples, and the isopure $^4$He.

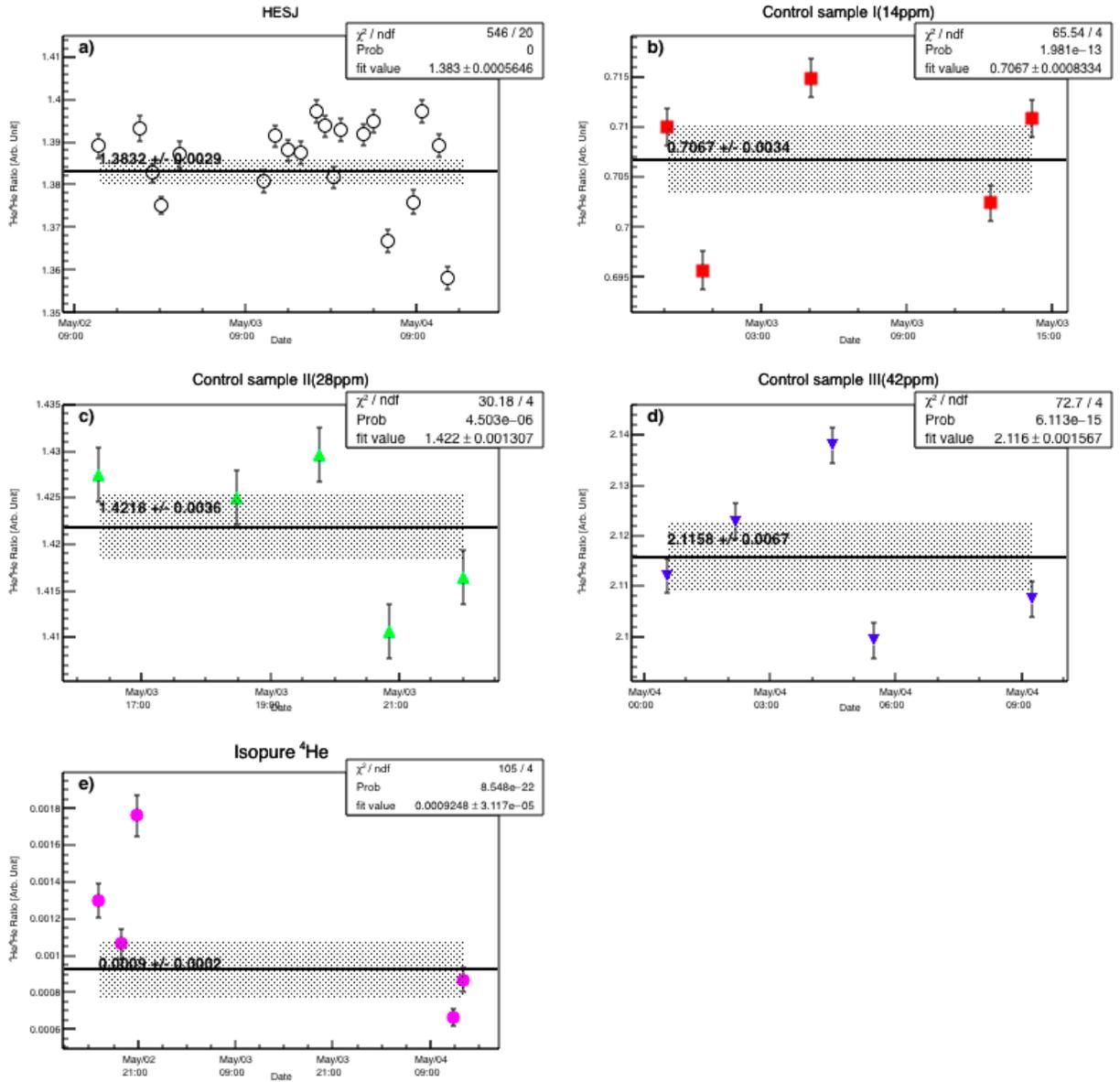

**Figure 6.** Time spectrum of measured ratios of $^3$He/$^4$He for the HESJ (a), control samples (b–d), and the isopure $^4$He (e) with a fit. The hatch shows 1 σ error, which were scaled by multiplying *S* in eq. (3).

The $^3$He/$^4$He values of control samples obtained by the MS are plotted in Figure 7, with values determined by gas expansion listed in Table 2 on the horizontal axis including a first-order polynomial fit function. The fit returned a reasonable $\chi^2$ value (2.8/2), which means that there was no $^3$He/$^4$He dependence in the measurements. Note that this analysis method was not affected by constant backgrounds. We estimate the upper limit of the constant background by the observed isopure $^4$He value, which is less than 0.1% of the HESJ. The HESJ gas measured in this study was newly taken from a distributed cylinder. The difference between the HESJ for this experiment and that stored in another gas container, which was taken from the original cylinder almost 20 years ago and used more than 460 times for daily calibration of the MS, was measured as 0.1 +/- 0.3 %. Thus, the effect of gas handling procedure and depletion of $^3$He/$^4$He of HESJ in the cylinder with time is negligibly small.

The $^3$He/$^4$He values of the MS can be converted into absolute $^3$He/$^4$He ratios with the fitting function. The $^3$He/$^4$He value of the HESJ measured by the MS is plotted on Figure 7 by a red band with its error corrected by the multiplicative factor $S$. The absolute $R_{HESJ}$ can be determined by the crossing point. Two uncertainties were taken into account to determine $R_{HESJ}$ by this method. One is an uncertainty caused by fitting of MS measurements. The uncertainty was evaluated by the errors of the fitting function. The other is an uncertainty caused by the fabrication of control samples, shown in Table 2. These uncertainties for the control samples were expected to correlate with each other. Thus, we treated them as to be fully correlated. We took following procedure to evaluate the uncertainties. First, fit with the central values of MS data shown in Figure 7. Next, the data points were shifted to the uncertainties caused by fabrication of the control samples. The data points shifted with upper and lower uncertainties of 1 σ were fitted to determine the uncertainties by the sample fabrications. Finally, the $R_{HESJ}$ was obtained as

$$R_{HESJ} = 27.36 \pm 0.08 \text{ (MS measurements)} \pm 0.08 \text{ (control samples fabrication) ppm}$$

$$= 27.36 \pm 0.11 \text{(combined) ppm,} \quad (4)$$

where the error designated as "MS measurements" means fitting error of the central value, and that as "control samples fabrication" means uncertainties with gas fabrication. The combined uncertainty was 0.40%.

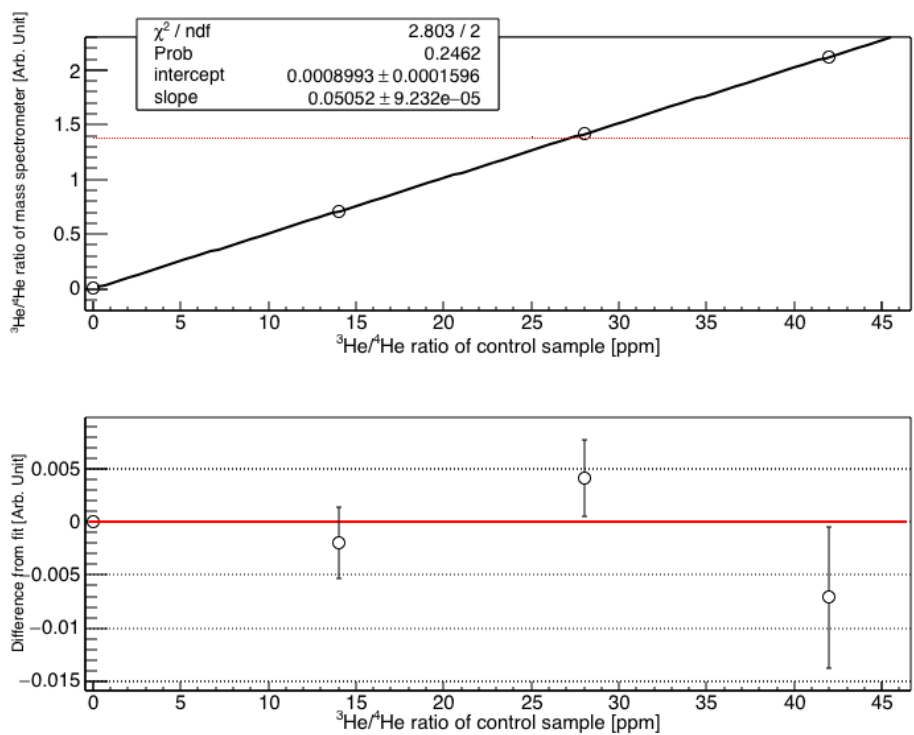

**Figure 7.** Plot and fit of the $^3$He/$^4$He ratio of the control samples: the expected values by the fabrication procedure vs. the values by measured MS (top), and its differences from the fit (bottom). The value for the HESJ is shown by a red dot band on the top graph.

## 4 $^3$He/$^4$He ratio of the atmosphere

The absolute value of the $^3$He/$^4$He ratio in the atmosphere, $R_a$, can be determined by that of the absolute $^3$He/$^4$He ratio of HESJ in the present work and the relative $^3$He/$^4$He ratio of HESJ to the atmospheric one, $R_{HESJ}/R_a$. The previous measurements of $R_{HESJ}/R_a$ values in Refs. (Matsuda et al. 2002; Lupton & Evans, 2004; Sano et al., 2008) are listed in Table 3.

**Table 3.** Measurements of $R_{HESJ}/R_a$.

| Publication year | $R_{HESJ}/R_a$ | 1 σ uncertainty | References |
|---|---|---|---|
| 2002 | 20.63 | 0.10 | Matsuda et al. (2002) |
| 2004 | 20.408 | 0.022 | Lupton & Evans (2004) |
| 2008 | 20.405 | 0.040 | Sano et al. (2008) |

We compiled the three measurements by taking the weighted mean, $R_{HESJ}/R_a$ was determined as

$$R_{HESJ}/R_a = 20.415 \pm 0.029, \qquad (5)$$

where we applied the scale factor of 1.5 caused by $\chi^2$/n.d.f. of 4.8/2 to reasonably evaluate the fitting result. By combining the $R_{HESJ}$ in eq.(4) and $R_{HESJ}/R_a$ in eq.(5), the $R_a$ was determined as

$$R_a = 1.3404 \pm 0.0056\ (R_{HESJ}) \pm 0.0019\ (R_{HESJ}/R_a)\ \text{ppm}$$

$$= 1.340 \pm 0.006\ \text{(combined) ppm}. \qquad (6)$$

The result, together with previous measurements (Mamyrin et al., 1970; Clarke et al., 1976; Davidson & Emerson, 1990; Sano et al., 1988; Hoffman & Nier, 1993), are listed in Table 4 and plotted in Figure 8. Note that the value for Davidson & Emerson (1990) was recalculated by the ratio of $^3$He to $^4$He content in the atmospheric air, where 7.27 ± 0.20 pptv for $^3$He and 5.2204 ± 0.0041 ppmv for $^4$He (Holland & Emerson, 1987), as 1.393 ± 0.38 ppm. The value of Sano et al. (1998) seems not to take into account systematic uncertainty of at least 0.6% for their calibration gases, however, we listed the value in the reference as it is because the effect is relatively small.

Table 4. Absolute $^3$He/$^4$He ratios of the atmospheric helium

| Sampling year | $^3$He/$^4$He ratio [ppm] | 1 σ uncertainty [ppm] | References |
|---|---|---|---|
| 1956 | 1.371 | 0.019 | Hoffman & Nier (1993) |
| 1969 | 1.399 | 0.013 | Mamyrin et al. (1970) |
| 1975 | 1.384 | 0.006 | Clarke et al. (1976) |
| 1988 | 1.343 | 0.013 | Sano et al. (1988) |
| 1988 | 1.393 | 0.038 | Davidson & Emerson (1990) |
| 2002, 2004, 2008 | 1.340 | 0.006 | This study |

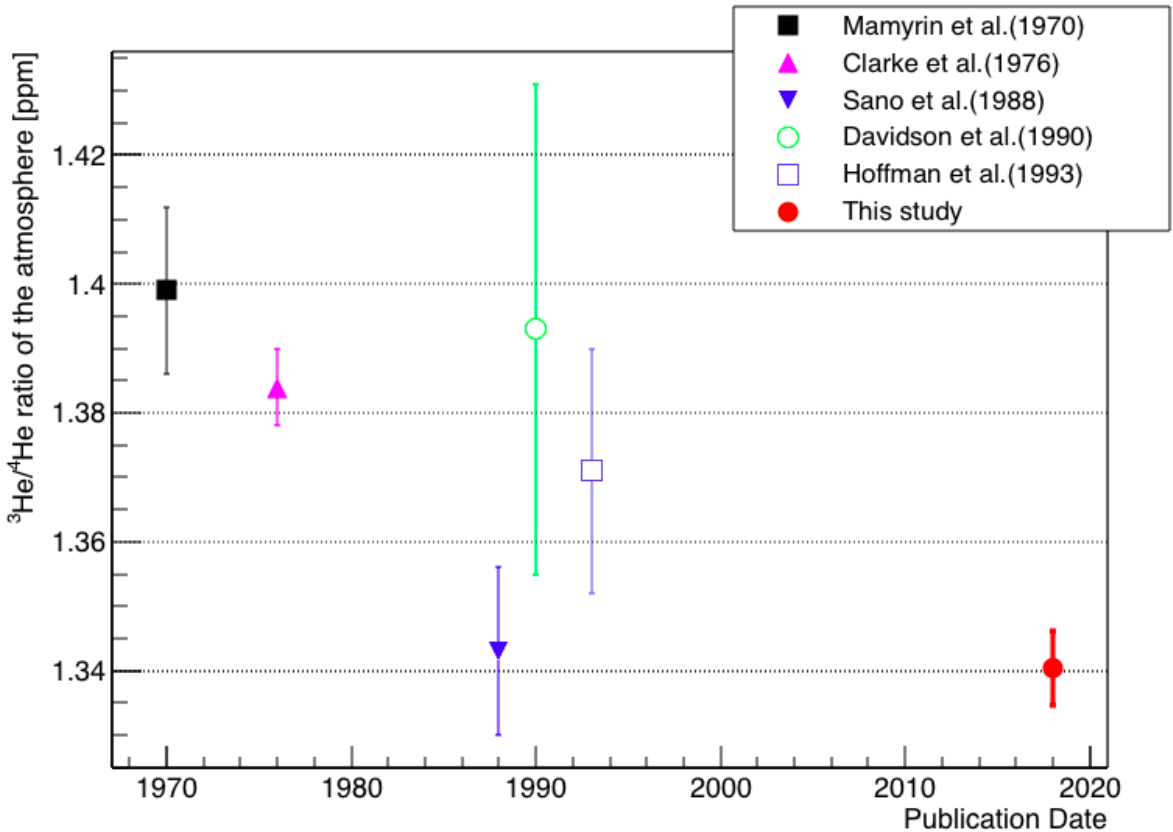

**Figure 8.** $^3$He/$^4$He ratios in the atmosphere (in units of ppm), present (red circle) and previously reported (Mamyrin et al., 1970; Clarke et al., 1976; Sano et al., 1988; Davidson & Emerson, 1990; Hoffman & Nier, 1993).

The present value is consistent with values reported by Sano et al. (1988), Davidson & Emerson (1990), and Hoffman & Nier (1993) in less than 1.6 σ, but different from those by Mamyrin et al. (1970) and Clarke et al. (1976) by 4.5 and 7.3 σ, respectively. The discrepancy must be examined as the latter two values have been widely accepted for $R_a$ value in many laboratories.

Our result relies on the $R_{HESHJ} / R_a$ value. Our compiled value, determined with accuracy of 0.14%, was dominated by two values: Lupton & Evans (2004) for their air standard taken in California, U.S. and Sano et al. (2008) taken in Tokyo, Japan. These two are consistent thus local heterogeneity of air $^3$He/$^4$He ratios cannot account for the apparent lower $R_a$ value by Sano et al. (1988) for the air sample collected in Japan than those by Mamyrin et al. (1970) and Clarke et al. (1976) for air taken in Russia and North America, respectively. Note that the value of Matsuda et al. (2002), was slightly different (2.2 σ) from the average, though it didn't make significant effect on the average due to its larger uncertainty than the others. Thus, $R_{HESHJ} / R_a$ is not possible to explain the difference.

The temporal variation of $^3$He/$^4$He ratio in the atmosphere, which was proposed by Sano et al. (1988; 2008), is a candidate of the cause of the difference. However, it was not supported by later studies (Lupton & Evans, 2013; Mabry et al., 2015), which gave the upper limit of the change as less than $4.5 \times 10^{-5}$/year and $4.2 \times 10^{-5}$/year (2 σ), respectively. Thus, even if the temporal variation is the case, it is too small to account for the difference. Another possibility is isotopic contaminations in isopure $^4$He or $^3$He gas. The isotopic purity of $^4$He was measured in this study, and the effect was estimated as less than 0.1% of that of HESJ. For $^3$He gas, we believed the isotopic composition provided by the manufacture, which is specified as more than 99.95%. Note that if any unexpected contamination in our $^3$He existed, the true $R_{HESJ}$ would be smaller, which would lead smaller $R_a$ value than 1.340 ppm. Thus, it cannot be the cause of the discrepancy.

In conclusion, we could not identify any reasonable reason of the discrepancy of our $R_a$ value from those determined by Mamyrin et al. (1970) and Clarke et al. (1976). The largest difference in the measurements is 4.4% between the present result and Mamyrin et al. (1970). Note that the difference of the absolute value of $R_a$ does not affect relative measurement of $^3$He/$^4$He ratio. However, the difference may be crucial for tritium-$^3$He dating of groundwater and seawater samples, which determines isolation time from the atmosphere through amount of tritium and $^3$He. The amount of $^3$He is usually determined by $^3$He/$^4$He ratio measurement. If the $^3$He/$^4$He ratios of samples were calibrated by $R_a$ value which was overestimated by 4.4%, the determined $^3$He amounts are needed to be corrected 4.4% smaller at maximum, which corresponds to 0.8 years shorter age. This correction is in the same order of magnitude to uncertainties of reported tritium-$^3$He ages under the ideal circumstance (Visser et al., 2014). However, as long as tritium concentration in the water sample is determined with the $^3$He in-growth method, in which newly produced $^3$He from tritium in the sample after complete degassing of originally-contained $^3$He and subsequent storage over a period (typically a couple of months) is measured via $^3$He/$^4$He analysis, the correction is also applied to tritium concentration to the same extent, resulting in no systematic error in tritium-$^3$He age as the age is a function of tritium/$^3$He ratio (Schlosser, 1992; Takaoka & Mizutani, 1987; Visser et al., 2014). On the other hand, if tritium in a water sample is determined by an independent method to that for $^3$He, such as beta counting of tritium decay, the possibility of systematic error on the tritium-$^3$He age should be assessed by a comparison with ages determined with other methods, such as CFC and SF$_6$ dating techniques.

In the case of the cosmogenic $^3$He-based surface exposure age determination, the $^3$He production rates had uncertainties of 4% or more (Niedermann, 2002). Thus, the difference of

4.4% is not significant for the age determinations. Note that in the case that a production rate was determined by a comparison of a known age of mineral/host rock and the amount of cosmogenic $^3$He measured by $^3$He/$^4$He analysis using a standard whose $^3$He/$^4$He ratios was calibrated to the atmospheric $^3$He/$^4$He, there is no effect on reported cosmogenic $^3$He age by the correction of atmospheric $^3$He/$^4$He value, where only absolute value of the production rate is needed to be changed.

## 5 Summary

The absolute value of the $^3$He/$^4$He ratio of the HESJ was measured in this work. We have fabricated control samples with $^3$He/$^4$He ratios of 14, 28, and 42 ppm, with uncertainties in the range of 0.25–0.38%, by using a gas handling system for a neutron lifetime experiment at the MLF BL05 in J-PARC (Arimoto et al., 2015; Nagakura et al., 2016). Their $^3$He/$^4$He values were compared with those of the HESJ using a magnetic-sector-type single-focusing noble gas MS with a double collector system, at the Department of Basic Science, of the University of Tokyo (Sumino et al., 2001). The $^3$He/$^4$He ratio of the HESJ was determined as 27.36 ± 0.11 ppm. This result can contribute to the improvement of the accuracy of neutron lifetime experiments (Arimoto et al., 2015; Nagakura et al., 2016; Mumm et al., 2016). With the present result and the averaged $R_{HESJ}$ / $R_a$ of 20.415 ± 0.029, the $^3$He/$^4$He in the atmospheric helium was determined as 1.340 ± 0.006 ppm, which is consistent with the recent IUPAC recommendation value, but not with some of other previous determinations (Clarke et al., 1976; Mamyrin et al., 1970) with discrepancy of ~4%. This is an important issue to be solved though we could not identify the reason of the discrepancy. A way to solve the problem is independent studies by other laboratories to fabricate their own control samples and compare with ours. Our control samples are possible to be provided.

## Appendix A Model calculation of gas diffusion

The gas mixture by diffusion in a tube can be described by the 1-dimension diffusion equation as

$$\frac{\partial u}{\partial t} = D \frac{\partial^2 u}{\partial^2 x} \qquad (7)$$

where $u$ is number density of atoms, $t$ is time, $x$ is position, and $D$ is diffusion constant. Assuming gas diffusion in a tube with length of $L$, the time constant, $\tau_d$, can be described as

$$\tau_d \cong \tau_1 = \frac{L^2}{\pi^2 D} \qquad (8)$$

where $\tau_1$ is the longest time constant in the solution. The diffusion constant of He gas of $1.013 \times 10^5$ Pa at 300 K with is $1.82 \times 10^{-4}$ m$^2$/s (Kestin et al., 1984). Assuming the diffusion constant $D$ is proportional to the pressure as the ideal gas, $D$ at 600 Pa, which is the condition for control sample fabrication, was calculated as $3.07 \times 10^{-2}$ m$^2$/s. We assume that diffusion in the 22-L buffer bottle is shorter enough because of larger diameter (⌀210 mm), resulting in the diffusion time dominated by the length of the tube (⌀7.4 mm). Assuming the tube length as 0.8 m from the sampling position to the buffer bottle, the $\tau_d$ is estimated as 200 s. The value is in the same order of magnitude as the measured diffusing time of 70 ± 3 s in Sec 2.3.

## Acknowledgments

This work was supported by JSPS KAKENHI, Grant Numbers JP16H02194 (to K.M.) and JP24654058 (to H.S.). The work at the Materials and Life Science Experimental Facility of the J-PARC was supported by S-type project of KEK (Proposal No. 2014S03). All data and fit results in Figure 5 can be found in the Supporting Information file.